\documentclass[prl,aps,twocolumn,superscriptaddress]{revtex4}
\usepackage{graphicx}
\usepackage{epstopdf}
\usepackage{latexsym}
\usepackage{amsmath}
\begin{document}

\title{Universal thermal and electrical transport near the\\ superconductor-metal quantum 
phase transition in 
nanowires}

\author{Adrian Del Maestro}
\affiliation{Department of Physics, Harvard University, Cambridge,
MA 02138}
\author{Bernd Rosenow}
\affiliation{Department of Physics, Harvard University, Cambridge,
MA 02138}

\author{Nayana Shah}
\affiliation{Department of Physics, University of Illinois at
Urbana-Champaign, 1110 W. Green St, Urbana, IL 61801}

\author{Subir Sachdev}
\affiliation{Department of Physics, Harvard University, Cambridge,
MA 02138}

\date{\today}

\begin{abstract}
We describe the thermal ($\kappa$) and electrical ($\sigma$)  conductivities of
quasi-one dimensional wires, across a quantum phase transition from
a superconductor to a metal induced by pairbreaking perturbations.
Fluctuation corrections to BCS theory motivate a field theory for quantum
criticality. We describe deviations in the Wiedemann-Franz ratio $\kappa/\sigma T$ (where $T$ is 
the temperature)
from the Lorenz number 
$(\pi^2 /3) (k_B /e)^2$, which can act as sensitive tests of the theory. We also describe the crossovers out of the quantum critical region
into the metallic and superconducting phases.
\end{abstract}
\maketitle

The Wiedemann-Franz law relates the low temperature ($T$) limit of the 
ratio $W \equiv \kappa/(\sigma T)$ 
of the thermal ($\kappa$) and electrical ($\sigma$) conductivities of metals to the universal 
Lorenz number $ L_0=
(\pi^2 /3) (k_B/e)^2$. This remarkable relationship is independent of the strength of the 
interactions 
between the electrons, relates macroscopic transport properties to fundamental constants of 
nature, and 
depends only upon the Fermi statistics and charge of the elementary quasiparticle 
excitations of the metal. It 
has been experimentally verified to high precision in a wide range of metals \cite{louis}, and realizes a 
sensitive 
macroscopic test of the quantum statistics of the charge carriers.

It is interesting to note the value of the Wiedemann-Franz ratio in some other important
strongly interacting quantum systems. In superconductors, which have low energy bosonic 
quasiparticle 
excitations, $\sigma$ is infinite for a range of $T>0$,
while $\kappa$ is finite in the presence of impurities \cite{smitha}, and so $W=0$. 
At quantum phase transitions described by relativistic field theories, such as the superfluid-insulator 
transition in the Bose
Hubbard model, the low energy excitations are strongly coupled and quasiparticles are not 
well defined; in 
such theories the conservation of the relativistic stress-energy tensor implies that 
$\kappa$ is infinite, and so $W=\infty$ \cite{vojta}. Li and Orignac \cite{edmond} computed $W$ in disordered Luttinger liquids, and found deviations from $L_0$, and found a non-zero universal value for $W$ at the metal-insulator transition for spinless fermions.

The present paper will focus on the quantum phase transition between a superconductor and a metal (a 
SMT). We will consider quasi-one dimensional nanowires with a large number of transverse channels (so that the electronic localization length is much larger than the mean free path ($\ell$)) which can model numerous recent 
experiments 
\cite{bezryadin-lau,liu-zadorozhny,lau-markovic,rogachev-bezryadin,boogaard,rogachev-bollinger,bollinger-roachev,chang,rogachev-wei}. We will describe universal 
deviations in the value of $W$ from $L_0$, which can serve as sensitive tests of the theory in future experiments. 

The mean-field theory for the SMT goes back to the early work \cite{ag} of Abrikosov and Gorkov (AG): in 
one of the earliest discussions of a quantum phase transition, they showed that a large enough 
concentration of magnetic impurities could induce a SMT at $T=0$. It has since been shown that such a 
theory applies in a large variety of situations with `pair-breaking' perturbations: anisotropic superconductors 
with non-magnetic impurities \cite{herbut},  lower-dimensional superconductors with magnetic fields 
oriented in a direction parallel to the Cooper pair motion \cite{lsv}, and $s$-wave superconductors with 
inhomogeneity in the strength of the attractive BCS interaction \cite{spivak-zyuzin}. Indeed, it is expected 
that pair-breaking is present in any experimentally realizable SMT at $T=0$: in the nanowire
experiments, explicit evidence for pair-breaking magnetic moments on the wire surface was presented recently by Rogachev {\em et al.} \cite{rogachev-wei}. 

Fluctuations about the AG theory have been considered \cite
{lsv,AL,MT} in the metallic state, and lead to the well-known Aslamazov-Larkin (AL), Maki-Thomson (MT) and Density of States (DoS) corrections to the conductivity. 
At the SMT, field-theoretic analyses \cite{swt,pv} 
show that the AG theory, along with the AL, MT and DoS corrections, is inadequate in spatial dimension $d \leq 2$, and additional 
self-interactions among Cooper pairs have to be included. Here, $d$ defines the dimensionality of the Cooper pair motion, while 
the metallic fermionic quasiparticles retain a three-dimensional character; therefore, the 
confining dimension, $R$, is larger than the inverse Fermi wavevector, but smaller than a superconducting 
coherence length or Cooper pair size, $\xi$. The behavior of $W$ has been considered in this field-theoretic framework 
\cite{pv}, and it was found that there were logarithmic corrections to the Lorenz number in $d=2$. Here we 
will examine the $d=1$ case in some detail: the transition is described by a strongly-coupled field theory of 
bosonic Cooper pairs, overdamped by their coupling to the fermionic quasiparticles. Remarkably, all 
important couplings between the bosons and the fermions scale to universal values, and consequently the 
Wiedemann-Franz ratio of this theory also approaches a universal constant which we compute in a $1/N$ expansion (the physical case is $N=1$)
\begin{equation}
W =  \left( \frac{k_B}{e} \right)^2 \left( 0.2820 + \frac{0.0376}{N} \right).
\label{resw}
\end{equation}
Our present computation of $W$ ignores the influence of disorder on quantum criticality, and this may require the clean limit $\xi \ll \ell$ \cite{caveat}.

\begin{figure}[t]
\centering \includegraphics[width=3.3in]{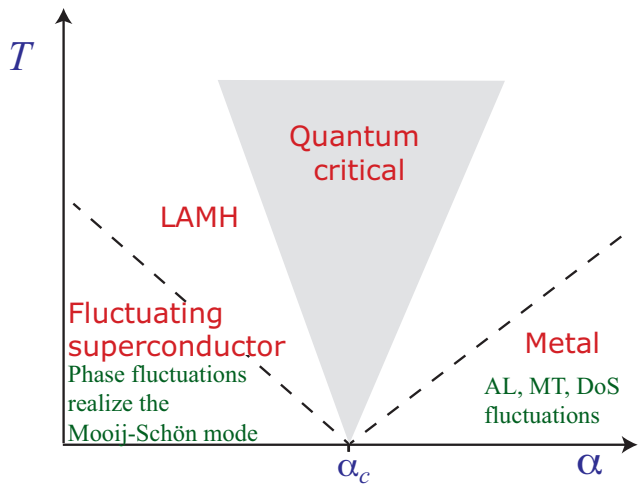} 
\caption{(Color online) Crossover phase diagram of the superconductor-metal transition in a quasi-one dimensional 
superconductor. The ``Metal'' is described by the perturbative theory of 
Ref.~\onlinecite{lsv}. The ``Quantum critical'' region is described by $\mathcal{S}$
and realizes our result for $W$ in Eq.~(\ref{resw}). The Mooij-Sch\"on mode is present
everywhere, but couples strongly to superconducting fluctuations only in the ``Fluctuating
superconductor'' regime, where it is described by Eq.~(\ref{phase}); note that $\mathcal{S}$
does {\em not\/} apply here. The dashed lines are crossover
boundaries which, by Eq.~(\ref{sscale}) occur at $T \sim |\alpha-\alpha_c|^{z \nu}$.} 
\label{phasediag}
\end{figure}

We also discuss the nature of the crossovers from this universal quantum critical physics to previously 
studied regimes at low $T$ about the superconducting and metallic phases: these are summarized in Fig.~\ref{phasediag}. 
On the metallic side, there is a crossover to a low $T$ regime described by the theory \cite{lsv} of AL+MT+DoS corrections in $d=1$. On the superconducting side, there is a regime of intermediate temperatures 
where the classical phase slip theory of Langer, Ambegaokar, McCumber, and Halperin (LAMH) applies \cite
{la,mh}, and eventually another crossover at still lower temperatures to a phase fluctuating regime 
whose description requires a non-linear $\sigma$-model of fermion pair fluctuations coupled to the superconducting
order \cite{andreev}. Here the phase fluctuations are essentially equivalent to a plasma 
charge oscillation, which in $d=1$ is the Mooij-Sch\"on mode \cite{ms}. A number of works 
\cite{zaikin,buchler,refael,oreg} have examined the destruction of superconductivity due to 
quantum phase slips in such a phase fluctuating regime: we maintain that the phase with phase slip 
proliferation in such models is an {\em insulator\/} at $T=0$, and so
such theories describe a superconductor-insulator quantum transition, and may be appropriate in 
inhomogenous systems. Our theory includes amplitude 
and phase fluctuations on an equal footing, along with strong damping from the fermionic modes, and describes the transition into a metallic phase at $T=0$: this is the case even though 
physics of the Mooij-Sch\"on mode is present in the ``fluctuating
superconductor'' regime in Fig.~\ref
{phasediag}.

It is useful to place our results in the context of recent microscopic computations in BCS theory \cite{lsv}
on the metallic side, with the pairbreaking parameter $\alpha$ larger 
than critical
$\alpha_c$ of the SMT. These results were obtain in the dirty limit ($\ell \ll \xi$), but we expect the same theory  (the action $\mathcal{S}$ below) to apply in the 
quantum critical regime 
in the both the clean and dirty limits (although there are distinctions in the ``fluctuating superconductor'' regime of Fig.~\ref{phasediag}). For the conductivity, these results are \cite{lsv}
\begin{eqnarray}
\sigma &=& \sigma_0 + \frac{e^2}{\hbar} \left(\frac{k_B T}{\hbar D} \right)^{-1/2} \left[ \frac{ \pi }
{12 \sqrt{2}}
\left( \frac{k_B T}{\hbar(\alpha - \alpha_c)} \right)^{5/2} \right]  \nonumber \\
&~&~~~~~~~+ \frac{e^2}{\hbar} \left(\frac{k_B T}{\hbar D} \right) \left[ c \frac{\hbar(\alpha - 
\alpha_c)}{k_B T} \right]  \label{lsv}
\end{eqnarray}
where $\sigma_0$ is a background metallic conductivity, $c$ is a non-universal constant, $D$ is the diffusion constant in the metal, 
and the remaining corrections from pairing fluctuations have
been written in the form of a power of $T$ times a factor within the square brackets which depends only 
upon
the ratio $\hbar(\alpha-\alpha_c)/k_B T$. This way of writing the results allows us to deduce the importance 
of the fluctuations corrections, in the renormalization group sense, to the SMT. The first square bracket 
represents the usual AL correction, and has a prefactor of a negative power of $T$, and so is a relevant 
perturbation; this is so even
though this correction vanishes as $T \rightarrow 0$. The second square bracket arises from the additional AL, MT and DoS corrections: the prefactor has no divergence as a power of $T$, and so this 
correction is formally
irrelevant at the SMT. Note, however, the complete second term has a finite limit as $T \rightarrow 0$, and so 
becomes
larger than the formally relevant AL term at sufficiently low $T$ in the metal.  We therefore identify the second 
term as {\em dangerously irrelevant\/} in critical phenomena parlance: {\em i.e.\/} important for the 
properties of the 
low $T$ metallic region, but can be safely neglected in the shaded quantum critical region of Fig.~\ref
{phasediag}. 

Armed with the above insights, we focus on the fluctuations associated with the usual AL correction. These have 
\cite{lsv} a Cooper pair propagator $(\widetilde{D} q^2 + |\omega| + \alpha)^{-1}$ at wavevector $q$ and imaginary 
frequency $\omega$ in the metal in both the clean and dirty limits. 
This motivates the quantum critical theory of Ref.~\onlinecite{swt} for a field $\Psi(x,
\tau)$ representing the local Cooper pair creation operator:
\begin{eqnarray}
\mathcal{S} &=& \int dx \Biggl[ d \tau \left( \widetilde{D} |\partial_x \Psi |^2 + \alpha |\Psi|^2 + \frac{u}{2} |\Psi|
^4 \right) \nonumber \\ &~&~~~~~~~~~~~~~
+ \int \frac{d \omega}{2 \pi} |\omega| |\Psi (x ,\omega)|^2 \Biggr] .
\end{eqnarray}
This theory will apply to quasi-one dimensional wires for $R < (\hbar \widetilde{D}/k_B T)^{1/2}$. In the dirty limit, $\ell \ll \xi$, we have $\widetilde{D} = D$, but the value of $\widetilde{D}$ is different in the clean case. All couplings in $\mathcal{S}$ are random functions of position;  in particular, randomness in $\alpha$ is expected to be relevant at the quantum critical point. We neglect this randomness in our quantitative results in the quantum critical region, and so they only apply above a $T=T_{\rm dis}$ which can be made arbitrarily small in the clean limit. 

From $\mathcal{S}$, we obtain\cite{swt}
the singular contribution to the conductivity in the vicinity of the quantum critical region
\begin{equation}
\sigma_{\rm sing} =  \frac{e^2}{\hbar} \left(\frac{k_B T}{\hbar \widetilde{D}} \right)^{-1/z} \Phi_\sigma \left
( \frac{\hbar(\alpha-\alpha_c)}{(k_B T)^{1/(z \nu)}} \right) \label{sscale}
\end{equation}
where $z$ is the dynamic critical exponent, $\nu$ is a correlation length exponent, and $\Phi_\sigma$ is a 
universal scaling function. In a Gaussian approximation, the Kubo formula yields $z=2$, $\nu=1/2$, and $\Phi_\sigma (y) = (\pi/(12 \sqrt
{2})) y^{-5/2}$, and so this result is in precise correspondence with the first term in Eq.~(\ref{lsv}). 
In the limit $T \ll (\alpha-\alpha_c)^{z \nu}$, we have already seen that this term is subdominant to the 
dangerously irrelevant second term in Eq.~(\ref{lsv}). However, moving into the quantum critical region where $ T \gg (\alpha-
\alpha_c)^{z \nu}$, the contribution from Eq.~(\ref{sscale}) dominates all other terms, and we have $
\sigma \sim T^{-1/z} \Phi_{\sigma} (0)$. The microscopic analysis of Ref.~\onlinecite{lsv} 
obtained $\sigma \sim T^{-2}$, which is valid only at $T$ large enough where $u$ can be neglected. Going beyond the Gaussian theory, the values of $z$ and $\nu$ have 
been determined in a $d=2-\epsilon$ expansion \cite{pankov,swt}, and also in quantum Monte Carlo 
simulations \cite{wts} with excellent agreement. 
Here, we have obtained these exponents in a theory with $N$ complex fields $\Psi$ directly in $d=1$ ,
and obtained to order $1/N$
\begin{equation}  
z = 2 - \frac{0.131}{N}~~~:~~~\nu = 1 - \frac{0.389}{N} ,
\end{equation}
to be compared with the Monte Carlo estimates of $z=1.97$ and $\nu=0.689$ \cite{wts}.
The value of $\Phi_\sigma (0)$ has a non-universal cutoff dependence associated with anomalous dimension 
$2-z$. Note that the results above for $\sigma$ in the metallic and quantum-critical regimes imply a non-monotonic $T$ dependence for $\alpha > \alpha_c$, possibly consistent with the observations of Ref.~\onlinecite{bollinger-roachev}.

Similar reasoning can be applied to the thermal conductivity $\kappa$, which can be computed 
from $\mathcal{S}$ using a separate Kubo formula \cite{moreno}. The scaling form analogous to Eq.~(\ref
{sscale}) is 
\begin{equation}
\kappa_{\rm sing} =  \frac{k_B^2 T}{\hbar} \left(\frac{k_B T}{\hbar \widetilde{D}} \right)^{-1/z} \Phi_\kappa \left( \frac{\hbar
(\alpha-\alpha_c)}{(k_B T)^{1/(z \nu)}} \right) \label{kscale}
\end{equation}
with $\Phi_\kappa$ another universal function. We have verified that the Gaussian prediction from $
\mathcal{S}$ again agrees with the perturbative AL contribution of the microscopic theory at low $T$. Our 
main result for the Wiedemann-Franz ratio in Eq.~(\ref{resw}) follows from $W = (k_B/e)^2 \Phi_\kappa 
(0)/\Phi_\sigma (0)$, as the nonuniversal  prefactor cancels out in ratio of these scaling functions; the $1/N$ 
expansion was carried out by generalizing the methods of Ref.~\onlinecite{csy}.


We now turn to an important conceptual issue:
the role of charge conservation and associated normal modes. From hydrodynamic arguments we know that 
a one-dimensional metal
or superconductor should support a gapless plasmon, or a Mooij-Sch\"on normal mode \cite{ms}. In $d=1$, 
this mode is gapless and disperses as $\omega \sim q \ln^{1/2} (1/(qR))$. In our theory for the quantum 
critical region, the Cooper pair field $\Psi$ carries charge $2e$ but only exhibits diffusive dynamics with $
\omega \sim \tilde{D} q^2$, and there is no Mooij-Sch\"on mode in the dynamics of the action $
\mathcal{S}$. 
The answer to this puzzle is contained in arguments made in Refs.~\onlinecite{ioffe-millis} and~\onlinecite{scs} on the role of conservation laws in the critical fluctuations of quantum transitions in metallic 
systems for which the order parameter is overdamped (as is the case here). These early works considered 
the onset of spin-density wave order in a metal; in the quantum critical region, the spin excitations 
consisted of diffusive paramagnons whose dynamics did not conserve total spin. However, Ioffe and Millis 
\cite{ioffe-millis} argued that the Ward identities associated with spin conservation only imposed significant 
constraints on the effective action at $\omega \gtrsim q$, and played little role in the $\omega \sim q^2$ 
regime important for the critical fluctuations. Essentially the same argument can be applied here: the 
Mooij-Sch\"on mode is present only at relatively high frequencies $\omega \sim q$, and the critical theory $
\mathcal{S}$ describes the overdamped Cooper pair modes in the distinct region of phase space with $
\omega \sim q^2$. The Mooij-Sch\"on fluctuations lead to oscillations in the local electrochemical potential, but these 
remain essentially decoupled from the critical modes described by $\mathcal{S}$ \cite{ioffe-millis,scs} (see however, 
Eq.~(\ref{phasym}) below).
It must be noted that the action $\mathcal{S}$ is {\em not\/} valid for $\omega \sim 
q$ and a complete description in terms
of the underlying fermions is necessary to obtain the proper dynamics, which will contain the Mooij-Sch\"on 
mode, as required.

Further insight into this issue is gained by lowering the temperature from the quantum 
critical regime into the ``fluctuating superconductor'' regime of Fig.~\ref{phasediag} for $\alpha < \alpha_c
$. When $k_B T  < (\alpha_c - \alpha)$ the action $\mathcal{S}$ does {\em not\/} apply for the smallest 
wavevectors and frequencies. The reasons for this are again analogous to arguments made for the spin-density-wave ordering transition in metals, as discussed in Ref.~\onlinecite{book}. For the latter case, it 
was argued that with the emergence of long-range spin density wave order, the low energy fermionic 
particle-hole excitations at the ordering wavevector were gapped out, and so the diffusive paramagnon 
action applied only for energies larger than this gap. At energies smaller than the gap, 
spin-waves with dispersion $\omega \sim q$ emerge, as a consequence of Ward identities associated with 
spin conservation.  In the superconducting case of interest here, there is no true long-range 
superconducting order at any $T>0$, but the order is disrupted primarily by `renormalized classical'  
thermal fluctuations of the phase, $\phi$ of the complex $\Psi$ field.
We can now assume that there is a local pairing amplitude in the fermion spectrum, analogous to 
the spin-density wave order. Charge conservation plays an important role in the effective action of $\phi$ 
fluctuations, and the `spin-waves' in this case are, of course, just the Mooij-Sch\"on excitations. The low 
energy effective action for $\phi$ cannot be obtained from $\mathcal{S}$; rather, we have to integrate the fermions out in the presence of a local pairing, as outlined above \cite{book}, and we expect an action 
\begin{eqnarray}
\mathcal{S}_\phi &=& \int \frac{dq}{2\pi} \int \frac{d \omega}{2 \pi} \frac{|A_\tau (q, \omega)|^2 }{4 
\ln (1/(qR))}  \nonumber \\
&+&    \int dx d\tau \Bigl [K_1 ( \partial_\tau \phi - 2 e A_\tau )^2 + K_2 (\partial _x \phi)^2 
\Bigr] 
\label{phase}
\end{eqnarray}
For the spin density wave case \cite{scs} it was found that $K_1 \sim (\alpha_c-\alpha)^{1/2}$ and
$K_2 \sim (\alpha_c - \alpha)$, and we expect similar behavior here.
We have explicitly included the action of the scalar 
potential $i A_\tau$ which mediates the Coulomb interaction in one dimension. The  normal modes of $
\mathcal{S}_\phi$ are the Mooij-Sch\"on oscillations which are now
identified with the fluctuations of the superconducting order. This should be contrasted from the situation in 
the shaded quantum critical region of Fig.~\ref{phasediag}, where the Mooij-Sch\"on oscillations were 
decoupled from the critical modes described by $\mathcal{S}$.

Actually, there is a weak coupling between the critical modes of $\Psi$ in $\mathcal{S}$ and the 
Mooij-Sch\"on mode due to particle-hole asymmetry (there is no 
analog of this phenomenon for the spin density wave case).
In general, the critical $\Psi$ action contains the perturbation \cite{swt}
\begin{equation}
\mathcal{S}_1 = \int dx d \tau \left[ \gamma \Psi^\ast \left( \frac{\partial}{\partial \tau} - 2ei A_\tau 
\right) \Psi \right]
\label{phasym}
\end{equation}
where $\gamma$ is proportional to the energy derivative of the density of states, and hence small.
Indeed, the same ratio of the pairing to Fermi energy, which justified the present quasi-one dimensional 
treatment of Cooper pairs but not electrons, also causes $\gamma$ to be small.
We examined the renormalization group fate of the perturbation $\mathcal{S}_1$ at the fixed point of $
\mathcal{S}$: we computed the two-loop flow of $\gamma$ in the $d=(2-\epsilon)$ expansion and found
\begin{equation}
\frac{d \gamma}{d \ell} = \frac{\epsilon^2}{100} \left( \pi^2 - 8 \right)\gamma.
\end{equation}
So $\gamma$ is relevant, but the scaling dimension is extremely small. Along with small bare value of $\gamma
$, such particle-hole asymmetric effects can justifiably be ignored in experimental applications.

The main experimentally testable results of this paper are: the crossover phase diagram in Fig.~\ref{phasediag}, the 
Wiedemann-Franz ratio in Eq.~(\ref{resw}) which applies in the shaded region, and the non-monotonic
$T$ dependence of $\sigma$ for $\alpha>\alpha_c$. 

We thank E.~Demler, B.~Halperin, G.~Rafael, and Y.~Oreg for useful discussions.
This research was supported by NSF grants DMR-0537077 and DMR-0605813, NSERC of
Canada (AD), and the Heisenberg program of DFG (BR).  Computing resources were provided by the Harvard Center for Nanoscale 
Systems, part of the National Nanotechnology Infrastructure Network.


\begin{thebibliography}{99}
\bibitem{caveat} A caveat for the clean limit is that, depending upon precise experimental conditions, the background, $T$-independent, contribution $\sigma_0$
 may be larger than $\sigma_{\rm sing}$ over the measured $T$ range. In this case, it should be kept in mind that the universal $W$ in Eq.~(\ref{resw}) of the theory $\mathcal{S}$ equals $\kappa_{\rm sing}/(T \sigma_{\rm sing})$.
\end{thebibliography}
\end{document}